\documentclass[12pt]{iopart}

\usepackage{color}
\usepackage{ulem}
\usepackage{graphicx}
\expandafter\let\csname equation*\endcsname\relax      
\expandafter\let\csname endequation*\endcsname\relax   
\usepackage{amsmath} 
\usepackage{amssymb} 
\usepackage{dcolumn}
\usepackage{bm}
\usepackage[utf8]{inputenc}
\usepackage[T1]{fontenc}
\usepackage{mathptmx}
\usepackage{siunitx}
\usepackage{hyperref}  
\begin{document}

\title[]{Validation of the plasma-wall self-organization model for density limit in ECRH-assisted start-up of Ohmic discharges on J-TEXT}

\author{Jiaxing Liu$^{1}$, Ping Zhu$^{1,2*}$, Dominique Franck Escande$^{3}$,
  Junli Zhang$^{1}$, Donghui Xia$^{1*}$, Yuhan Wang$^{1}$, Jiaming Wang$^{1}$,
  Qinghu Yang$^{1}$, Jiangang Fang$^{1}$, Li Gao$^{1}$, Zhifeng Cheng$^{1}$,
  Zhipeng Chen$^{1}$, Zhoujun Yang$^{1}$, Zhongyong Chen$^{1}$, Yonghua
  Ding$^{1}$, Yuan Pan$^{1}$ and the J-TEXT team$\footnote{See the author list of "N. Wang et al 2022 Advances in physics and applications of 3D magnetic perturbations on
    the J-TEXT tokamak, Nucl. Fusion 62 042016"}$}
\address{
	{\small $^1$International Joint Research Laboratory of Magnetic Confinement Fusion 
		and Plasma Physics, State Key Laboratory of Advanced Electromagnetic 
		Engineering and Technology, School of Electrical and Electronic 
		Engineering, Huazhong University of Science and Technology, Wuhan, 
		430074, China.} \\
	{\small $^2$Department of Engineering Physics, University of Wisconsin-Madison, Madison, Wisconsin, 53706, United States of America.}\\
	{\small $^3$Aix-Marseille Universit$\acute{\text{e}}$, CNRS, PIIM, UMR 7345, Marseille, France.}
}
\newcommand{\eadss}[1]{\address{E-mails: #1}}
\eadss{zhup@hust.edu.cn, xiadh@hust.edu.cn}
\vspace{10pt}
\begin{indented}
\item[]\today
\end{indented}

\begin{abstract}
  A recently developed plasma-wall self-organization (PWSO) model predicts a
  significantly enhanced density limit, which may be attainable in tokamaks with
  ECRH-assisted ohmic startup and sufficiently high initial neutral density.
  Experiments have been conducted on J-TEXT to validate such a density limit
  scenario based on this model. Experimental results demonstrate that increasing
  the pre-filled gas pressure or ECRH power during the startup phase can
  effectively enhance plasma purity and raise the density limit at the flat-top.
  Despite the dominant carbon fraction in the wall material, some discharges
  approach the edge of the density-free regime of the 1D model of PWSO.
\end{abstract}

%
\vspace{2pc}

\section{Introduction}
A high plasma density in a fusion device is crucial for fulfilling the Lawson
criterion for ignition. However, there exists an upper limit to this density,
above which a disruption may occur, leading to the termination of the
discharge. Since this density limit sets a stringent bound on the stable
performance and ignition regimes of a tokamak, its scaling and the associated
physics has been a subject of primary interests.

For decades, people used the empirical scaling law for the tokamak density limit
$\displaystyle
n_{\text{\tiny GW}}\left(\SI{}{m}^{-3}\right)=\frac{I_\text{p}\left(\SI{}{MA}\right)}{\pi
  a\left(\SI{}{m}\right)^2}\times 10^{20}$, where $I_\text{p}$ is the plasma
current and $a$ is the minor radius \cite{Greenwald_1988,Greenwald_2002}.
 Recently, a power balance model  {considering radiation} introduced a modified scaling
 for the density limit, $(I_\text{p}P/a^4 )^{(4/9)}$,
 where  $P$ is the heating power \cite{Zanca_2017,Zanca_2019,Zanca_2022}.
 {This radiative scaling} is in better agreement with the tokamak and reversed field pinch (RFP)
 experimental databases. The same model yields a  good scaling for the
 stellarator too \cite{Zanca_2017,Zanca_2022,Fuchert_2018}.
  {The primary factors influencing these power balance limits stem from impurity
 radiation, which is largely controlled by plasma-wall interactions
 \cite{Zanca_2017,Zanca_2019,Zanca_2022}.}  {This radiation affects the amount of
 heat reaching the limiter/divertor targets, subsequently determining the
 temperature in the target region. Additionally, the target region temperature
 significantly impacts impurity production, which in turn influences impurity
 radiation. This feedback mechanism forms the foundation of the recently
 proposed plasma-wall self-organization theory \cite{escande2022}.} This
 self-organization mechanism yields a steady level of impurity radiation only
 when the plasma density is below a certain  {radiative} limit
 \cite{escande2022}.
  {A higher density limit is reached in stellarator when the start-up is
 performed by using higher ECRH power \cite{Klinge2019,Wolf2019}. This higher density limit might be due to
 their mode of breakdown at start-up phase: the massive use of
 ECRH power with high neutral density producing less impurities
 \cite{Zanca_2017,escande2022}.} Furthermore, in W7-X the effective plasma charge
 {$Z_\text{eff}$} decreases with ECRH power \cite{Pavone_2019}. This suggests
 that increasing progressively and simultaneously the ECRH power and the initial
 neutral density could also decrease the initial production of impurities in tokamaks
 \cite{escande2022}. And this may increase the above radiative density limit.

   In order to investigate the impact of start-up conditions {,
     including pre-filled neutral gas pressure and ECRH power,} on the flat-top
   density limit in ECRH assisted ohmic discharges, a series of experiments are
   conducted on the J-TEXT tokamak. The experimental results indicate that
   increasing either the pre-filled gas pressure or the ECRH power in the
   start-up phase leads to a reduction in impurity radiation, an increase in the
   boundary electron temperature during the flat-top phase, and an enhancement
   of the density limit in most of the shots. Besides,  {the
     density limit is calculated using PWSO 0D and 1D model with parameters of
     J-TEXT}. The results demonstrated a general agreement with the experimental
   data under certain parameter assumptions.

   The remainder of this paper is organized as follows: Section 2
   introduces the experimental set-up. Section 3 details the experimental
   methods and presents the measurement results. Section 4 offers a comparison
   and analysis of the experimental and calculated results. Lastly, Section 5
   concludes with a summary and discussion.

   \section{Experimental set-up}
   The J-TEXT tokamak is a medium sized iron-core tokamak with a major radius
$R_0 = \SI{1.05} {m}$, minor radius $a=25-\SI{29}{cm}$, and a
silicon-carbide-coated graphite limiter \cite{Liang_2019}. The typical J-TEXT
discharge in the limiter configuration is performed with a toroidal magnetic
field $B_\text{t}$ of $\sim \SI{2.0}{T}$, a plasma current $I_\text{p}$ of
$\sim\SI{200}{kA}$, plasma density
$n_\text{e}$ of $1-7\times10^{19}\SI{}{m}^{-3}$, and an electron temperature
$T_\text{e}$ of  {$\sim\SI{1}{keV}$} \cite{Liang_2019}. The Ohmic
discharges begin with a reversed current closed at $t=\SI{0.0}{ms}$ to increase the rate of
magnetic flux change. Then the capacitors bank, including the ionization
capacitors, discharges to the ohmic coils to induce a toroidal electric field,
ionizing the gas within the vacuum chamber. Following this, the capacitors for
rapid build up of plasma current are turned on.

The material of limiter targets in J-TEXT is carbon whose chemical sputtering by
hydrogen is not negligible in comparison to the physical sputtering
\cite{Stangeby2000}. The role of silicon carbide coating is to reduce the
sputtering of carbon. In the following calculation concerning the sputtering function,
its effects are neglected since the impurity radiation on J-TEXT
primarily originates from carbon. The limiter target can absorb a large amount
of gas in high electron density discharges, thus cleaning discharges have to
{be carried} out during the experiments. Since non-standard start-up settings
are used for this study, a noticeable amount of discharges fail or disrupt. Gas
puffing can be applied to inject hydrogen with various pre-filled gas pressure
and gas puffing rate.
The ECRH system on J-TEXT was installed in 2019 \cite{Xia2018}.  Generated by a
gyrotron, the maximum output power is $\SI{500} {kW}$ and the frequency is
\SI{105}{GHz}. The toroidal magnetic field range of the ECRH system operation is
$\SI{1.7}{T} \le {B_\text{t}}\le \SI{2}{T}$. The ECRH auxiliary heating system
operates at the second harmonic frequency with X mode \cite{zhang2020,Zhang2023}.

The primary diagnostics for this experiment include the photo-diode array (PDA),
Langmuir probe, polarimeter-interferometer (Polaris) system, and vacuum
gauge. The arrangement of these diagnostic systems, as viewed from the top, is
depicted in Fig.~$\ref{fig:J-TEXTtopview}$. The PDA array, responsible for
measuring $\text{C}_\text{III}$ and hydrogen-alpha radiations on the high field side, comprises 18 channels installed at the top of port 10. A Langmuir probe array
at port 13 measures the electron temperature around the top limiter.
The Polaris system is capable of measuring 17-channel line-averaged plasma
electron density at different major radii \cite{YHWANG_2022}.
The vacuum gauge quantifies the pre-filled gas pressure within the vacuum vessel.

\section{Experimental results}
\subsection{Baseline discharge}
A pure Ohmic start-up baseline discharge (\#1082483) is shown in
Fig.~\ref{fig:481_483_014}.
The discharge starts at \SI{0.0}{s}, with a toroidal magnetic field of
$\SI{1.875}{T}$ and a plasma current of $\SI{120}{kA}$.
Hydrogen gas is injected after an electric pulse is sent to the gas puffing
system, resulting in a neutral gas pressure of approximately $\SI{3}{mPa}$ at
$t=\SI{0}{s}$. During the current plateau period, gas injection continues until
the plasma density limit of $0.7 n_\text{G}$ is reached.
At this point, the plasma density starts to decrease and then rapidly drops to
zero, without any prior manifestation of MHD activities. More experimental details
can be found in recent reports \cite{zhang2020,Zhang2023}.

\subsection{Density limit parameter dependence}
  Many discharges are performed with varying pre-filled gas pressure and
  start-up ECRH power as shown in Fig.~\ref{fig:moredata}.
   {These} discharges are analyzed to obtain the dependence of
  density limit $n_{\text{limit}}$ on key parameters including the target region temperature $T_{\text{t}}$ and the
  $\text{C}_\text{III}$ radiation power.  {These experiments were
    completed on two experimental days, and the final density limit values
    achieved under identical input conditions may differ due to varying device setup states,
  such as the wall state, on different experimental days. However, this  does not
  affect the overall law of the experimental results.} The discharges shown in
  Fig.~\ref{fig:nlimit_CIII} and 
  Fig.~\ref{fig:nlimit_Te} have identical plasma current, toroidal magnetic
  field but differ in start-up conditions, either in terms of pre-filled gas
  pressure or ECRH power. The  {plasma temperature $T_\text{t}$
    values around the limiter target}
  and impurity radiation power $R_{\text{C}_\text{III}}$ in
  Fig.~\ref{fig:nlimit_CIII} and Fig.~\ref{fig:nlimit_Te} represent their average
  values over the time interval $\left[\SI{200}{ms},\SI{250}{ms}\right]$, during
  which they are relatively stable. The $\text{C}_\text{III}$ radiation and
  {target region} temperature at the current plateau phase can be adjusted by
  changing the start-up condition. The experimental data indicates that lower
  $\text{C}_\text{III}$ radiation and higher target region temperature generally
  lead to a higher density limit, as shown in Fig.~\ref{fig:nlimit_CIII} and
  Fig.~\ref{fig:nlimit_Te}. Subsequently, two typical examples of varying gas
  pressure and ECRH power will be discussed in detail. 

  \subsection{ {Typical discharge for changing gas pressure}}
  Discharge \#1083014 is {performed} with a higher pre-filled gas pressure than
  the baseline discharge. The neutral gas pressure $P_\text{neu}$ at
  $t=\SI{0}{s}$ is approximately three times higher as shown in
  Fig.~\ref{fig:481_483_014}. The gas pressure of shot \#1083014 remains
  higher than that of shot \#1082483 during current ramping phase
  $\left(\text{Fig.}~\ref{fig:481_483_014}\right)$. The $\text{C}_\text{III}$ radiation power
  initially oscillates and eventually approaches a steady level over time in
  both discharges. The plasma target region temperature is higher in the
  discharge with higher gas pressure after the radiation power stabilizes.
  During the current plateau 
  phase, hydrogen gas is injected at the same rate until the density limit
  disruption is triggered. A higher density limit of $0.745 n_\text{G}$ is achieved in the discharge
  with higher pre-filled gas pressure.  

\subsection{ {Typical discharge for changing ECRH power}}
 Another discharge, \#1082481, is performed with the application of ECRH over
 the time interval $\left[\SI{-25}{ms}, \SI{20}{ms}\right]$. The injected ECRH
 power is approximately $\SI{230}{kW}$. Concurrently, the hydrogen gas is
 injected at the same rate with the baseline discharge until the density limit
 is reached. The injected ECRH power before $t=\SI{0}{s}$ ionizes the pre-filled
 gas and forms the so-called pre-plasma. The $\text{C}_\text{III}$ radiation
 power oscillates and reaches a lower steady state than that of the baseline
 discharge. During the start-up phase, the temperature oscillation amplitude is
 slightly larger than that of the baseline discharge. And the steady target
 region temperature  remains consistently higher. During the current plateau
 phase, the plasma density in this discharge increases at nearly the same rate
 as that of the baseline discharge but for an extended duration, ultimately
 reaching a higher density limit of $0.756 n_\text{G}$.

 \section{Theory and experiment comparison}
 We further compare the predictions from the PWSO 0D and 1D
  models with the J-TEXT experimental results presented in the previous section.
\subsection{PWSO 0D model and comparison}
The basic idea of the PWSO theory ({section 4.1 of \cite{escande2022}}) is that
the existence of a time delay in the feedback loop relating impurity radiation
and impurity production on divertor/limiter plates yields a delay equation in the 0D model
\begin{equation}
	\label{equ:0Diter}
	R+ =\alpha(P-R)
\end{equation}
 where $P$ is the total input power
 to plasma, $R$ the total radiated power, and $R+$
 the delayed radiation power during the next
  cycle of the feedback loop.

 The coefficient $\alpha$ quantifies the radiation power $R+$
  generated by the impurity produced from the plasma-wall interaction that is
  proportional to the deposition of the outflow power $\left( P-R \right)$ onto
  the wall targets, which is modeled as \cite{escande2022}
\begin{equation}
  \label{eq:alpha}
	\alpha=\frac{f \lambda}{a D_{\perp} T_\text{t}} I\left(T_\text{t}\right) \int_{0}^{a} r n(r) \operatorname{Rad}[T(r)] \mathrm{d} r
\end{equation}
 where $a$ is the plasma radius, $D_\perp$ is the perpendicular
  diffusion coefficient, $T_\text{t}$ is the  {plasma temperature at the
    target plate location}, $f$ is the
  fraction of the sputtered atoms that reach the main plasma
  and become ionized at a distance $\lambda$  inwards from the
  plasma target location, Rad$\left[T\left( r \right) \right]$
    is the impurity radiation rate coefficient, and $I\left(T_\text{t}\right)$
    is  {an average of} the yield function of carbon $Y\left(E\right)$ over the
    energies of the impinging particles 
    \begin{equation}
    \label{eq:ITt}
    I\left(T_\text{t}\right)=\sqrt{\frac{m}{2\pi T_\text{t}}} \int_{0}^{\infty} Y\left(\frac{m v^{2}}{2}+\gamma T_\text{t}\right) \exp \frac{-m v^{2}}{2T_\text{t}} \mathrm{~d} v
  \end{equation}
  where  {$\gamma$ is the total energy transmission coefficient
    \cite{Stangeby2000}, $\gamma T_t$ is a measure of the Debye shield length,
    and} $m$ is the ion mass.   
  The fixed point of Eq.$\left( \ref{equ:0Diter} \right)$ $R= R+$
  corresponds to the plasma-wall self-organization equilibrium.
  The plasma-wall system is unstable for $\alpha>1$ as
  predicted from Eq.~$\left(\ref{equ:0Diter}\right)$.  {So the condition that the
  threshold $\alpha=1$ establishes} a radiation density limit
  \begin{equation}
    n_{c}=\frac{2 D_{\perp}}{f \lambda \operatorname{Rad}[T(r)]} \frac{T_\text{t}}{I\left(T_\text{t}\right) a}
    \label{equ:0dnlim}
  \end{equation}
which can be reached for a
ratio of total radiated power to total input power as low as 1/2
\cite{escande2022}.
   {If there is a detachment,} the plasma temperature at the
   plates decreases  {as shown in Fig.~$\ref{fig:alpha-Tt}$},
   which makes $\alpha$ to vanish, since sputtering does too. This pushes the
   radiative density limit to very high values, especially when physical
   sputtering dominates the contribution to radiation. There are two basins of
   PWSO at the flat top of plasma current. The usual one is the regime of
   density limit corresponding to the higher temperatures of targets, whereas
   the other is the regime of density freedom corresponding to the lower
   temperatures of  targets, in particular in machines where the target plates
   are made of high-Z materials. 
  
  The projectile particles in our experiments
  are deuterons and the target material is carbon.
  The yield function consists of two parts for the physical and chemical
  sputtering contributions. The interpolating functions for $Y_\text{phy}(E)$ at
  normal incidence are provided in the Eq.~$\left(15\right)$ of
  \cite{YAMAMURA1996} for physical sputtering
  \begin{equation}
    \label{eq:yphy}
    Y_\text{phy}(E)=0.042 \frac{Q\left(Z_{2}\right) \alpha^{*}\left(M_{2} /
        M_{1}\right)}{U_{\text{s}}}\frac{S_{\text{n}}(E)}{1+\Gamma k_{\text{e}} \epsilon^{0.3}}
    \times{\left[1-\sqrt{\frac{E_{\text{th}}}{E}}\right]^{s} }
  \end{equation}
   where the parameters  on the
    right hand of Eq.~$\left(\ref{eq:yphy}\right)$ can also be found.
    Here the numerical coefficient $0.042$ is in unit of {\r
      A}$^{-2}$, $Z_1$ and $Z_2$ are the atomic numbers, $M_1$ and $M_2$ are the
    masses of the projectile and the target atoms, respectively, $S_\text{n}$ is
    the reduced nuclear stopping cross section, $U_\text{s}$ is the surface
    binding energy of the target solid, $k_\text{e}$ is the Lindhard electronic
    stopping coefficient, $E$ is the projectile energy, $E_\text{th}$ is the
    threshold energy for sputtering, $\epsilon$ is the reduced energy
    $E\frac{M_2}{M_1+M_2}\frac{a_L}{Z_1Z_2e^2}$, the $\Gamma$ factor  has the
    form $W\left( Z_2 \right)/\left(1+\left( M_1/7\right)^3 \right)$, and $W$
    and $Q$ are dimensionless  {fitting} coefficients. The yield function
    $Y_\text{che}\left(E\right)$ for chemical sputtering is obtained from
    fitting the available experimental data in Fig.~3.8 of \cite{Stangeby2000}
    and Fig.~6 of \cite{Joachim2007} {using the least squares method} 
     \begin{equation}
       \label{eq:yche}
     Y_\text{l,che}\left( E \right)=\left\{
    \begin{array}{l}
     a_1E_\text{l}^2+b_1E_\text{l}+c_1 \quad \qquad \qquad E_\text{l}<{E_{\text{l,inter}}}\\
      a_2E_\text{l}+b_2\qquad\qquad\qquad\qquad E_\text{l}\ge{E_{\text{l,inter}}}
    \end{array}
  \right.
\end{equation}
 {where $E_\text{l}=\log_{10}\left( E \right), Y_\text{l,che}=\log_{10}Y_{\text{che}}$,
$Y_{\text{che}}$ represents the chemical sputtering yield function. $a_1, b_1, c_1, a_2, b_2$,
and $E_{\text{l,inter}}$ are the fitting coefficients. The values of these
coefficients can be found in \ref{app:fitting}, along with a plot illustrating
the fitted chemical sputtering yield function and the original data points}. For
our experiments on J-TEXT, the minor radius $a = \SI{0.265}{m}$, and the impurity radiation rate
$\operatorname{Rad}\left[T\left(r\right)\right]$ is assumed to be a constant
{value} $10^{-32} \SI{}{Wm}^{-3}$  {\cite{Zanca_2019,Stangeby2000}}. We further assume that  {the perpendicular
diffusion coefficient $D_\perp$ of target impurities is $0.01\SI{}{m}^2
\SI{}{s}^{-1}$, and one percent of the sputtered atoms penetrate the main
plasma}, undergoing ionization at a distance $\lambda=\SI{0.01}{m}$ away from the
target. The maximum energy carried by the projectile particle is assumed to be
$\SI{5000}{eV}$, which is needed in the integral operation of
Eq.~$\left(\ref{eq:ITt}\right)$. The relation between the density limit and the
temperature of target for our experiments as predicted by the PWSO 0D model is
thus calculated and shown in Fig.~\ref{fig:denlim-Tt}.

The PWSO 0D model predicts a {density-free} regime
{($T_\text{t} \lesssim \SI{2}{eV}$)} and a {density-limit}
regime ($T_\text{t} \gtrsim \SI{2}{eV}$). In the {density-free
  (density-limit)} regime, the density limit increases with the
{decrease (increase)} of the target region
{plasma} temperature. J-TEXT experimental {results
  are located} in the {density-limit} regime,
{which qualitatively agree with the PWSO 0D model predictions.}
\subsection{PWSO 1D model and comparison}
 {A more} detailed evolution of the radiation  {power} and
 the temperature towards the PWSO equilibrium
 profiles can be obtained from the PWSO 1D model  {(see appendix C of \cite{escande2022})}. 

    The impurity density and the plasma temperature evolution can be determined from the
    following 1D transport equations
    \begin{equation}
      \begin{array}{rl}
        &\partial_{t} n_{i}-D \partial_{x}^{2} n_{i}=C_{i}\left[\partial_{x} T\left(r_{\text {LCFS }}, t-\tau_{\text {delay }}\right)+T_{\text {loss }}^{\prime}\right] \delta(x-a+\lambda)~,
      \end{array}
      \label{equ:nitran}
    \end{equation}
    \begin{equation}
      n \partial_{t} T-K \partial_{x}^{2} T=C_T T^{3/2}+P_{\text {add }}-n n_{i} \operatorname{Rad}(T)~,
      \label{equ:Ttran}
    \end{equation}
    where $C_{i}=-\frac{f a K I\left(T_{\text{t}}\right)}{(a-\lambda) T_{\text{t}}}$ represents 
    the plasma-wall interaction, $K$ is a uniform diffusion coefficient, 
    $C_{T}=\frac{E_{0}^{2}}{\eta(T) T^{3 / 2}} \simeq 6.510^{2} \frac{E_{0}^{2}}{Z}$, 
    with $E_0$ the electric field corresponding
    to the loop voltage, $\eta(T)$ is the transverse Spitzer
    resistivity, and $P_{\text{add}}$ is the additional power
    density. The above parameters of J-TEXT are used to obtain the following
    results and  {the resistivity $\eta$ is considered to be a constant}. Applying
    the following initial and boundary conditions
    \begin{equation}
  \begin{array}{ll}
    n_{i}(x=a)=0 ; & \left.\frac{\partial n_{i}}{\partial x}\right|_{x=0}=0 \\
    T(x=a)=T_{0} ; & \left.\frac{\partial T}{\partial x}\right|_{x=0}=0 \\
    n_{i}(t=0)=0 ; & T(t=0)=T_{0}
  \end{array}  
\end{equation}
a relation between the density limit and the target  {region} temperature has
been obtained and compared with the experimental data as shown in
Fig.~\ref{fig:denlim-Tt}.

The PWSO 1D model also predicts the existence of a density-limit basin and a
{density-free} basin. The J-TEXT  experimental data are located in the density
limit basin,  as in the case of the 0D model. Within the density-limit basin,
altering the start-up condition, such as increasing the pre-filled gas pressure
and ECRH power at the start-up phase,  {should} result in lower
$\text{C}_\text{III}$ radiation power, indicating a cleaner plasma, which leads
to a higher density limit.  {The} higher target region plasma temperature is
likely  a consequence of the higher power flux in the SOL, due to lower
radiation inside the plasma. Additionally, the target region plasma temperature
for the transition between the two basins is predicted to be higher in the 1D
model ($T_\text{t}\sim\SI{6}{eV}$), thus allowing a more accessible density-free
basin, which may be achieved in future experiments.

The lower $\text{C}_\text{III}$ radiation power when using ECRH at sart-up
confirms the improvement of purity expected in \cite{escande2022} due to a
central breakdown of the discharge. 
\section{Summary}
In this work, the density limits predicted from the 0D and 1D PWSO models are
compared and validated with the J-TEXT experimental data, which are located in
the density-limit basin and demonstrate quantitative agreement with the PWSO
model predictions under some parameter assumptions. Both theory and experiment
suggest that a higher density limit corresponds to a lower impurity radiation
state, which could be reached through increasing either the ECRH power or the
pre-filled gas pressure at the start-up phase.
In fact, due to the easy absorption of gases by carbon targets and other
experimental conditions, it is difficult to start a discharge out of J-TEXT's
usual start-up conditions. So, experimentally, increasing either the ECRH power
or the pre-filled gas pressure at the start-up phase was not always efficient in
changing the start-up conditions. The fact that the experimental points are at
the edge of the density-free regime of the 1D model indicates the possibility of
reaching it by further experimental improvements in J-TEXT, despite of its
graphite targets. In contrast, experiments in a tokamak with tungsten targets
might enter deeply the density-free regime. Furthermore, metallic walls should
enable scans of initial neutral gas density and of ECRH power with less failures
and disruptions at start-up than with carbon ones.

In future, we plan to carry out more experiments to explore whether or how the
J-TEXT tokamak can operate in the density-free basin predicted by the PWSO
model. In the meantime, this model is being implemented in  {a} more complete
transport simulation code to predict and interpret the experimental process in
a more realistic way.
\section{Acknowledgment}
This work is supported by the National MCF Energy R\&D Program of China under
Grant Nos.~2019YFE03050004 and 2018YFE0310300, the National Natural Science
Foundation of China Grant No.~51821005, and the U.S. Department of Energy Grant
Nos.~DE-FG02-86ER53218 and DE-SC0018001. The computing work in this paper is
supported by the Public Service Platform of High Performance Computing by
Network and Computing Center of HUST.
\appendix
\section{The fitted chemical sputtering yield function}
\label{app:fitting}
The data used to fit Eq.~\ref{eq:yche} are from Fig.~3.8a of \cite{Stangeby2000} and Fig.~6 of
\cite{Joachim2007}. The values of fitting coefficients are as follows:
$a_1=-0.21716823, b_1= 1.49621640, c_1=-3.10869745, a_2=-0.55766667,
b_2=-0.88506667$, and $E_{\text{l,inter}}=1.24709069$. And
Fig.~\ref{fig:chem_sput} shows the plot of original data used and the fitting
function. 

\makeatletter
\renewcommand{\thesection}{\@arabic\c@section}
\makeatother
\renewcommand{\thefigure}{\arabic{figure}}

\section*{References}
\bibliographystyle{unsrt}
\bibliography{references}

\newpage
\begin{figure}[htbp]
	\centering
	\includegraphics[width=\linewidth]{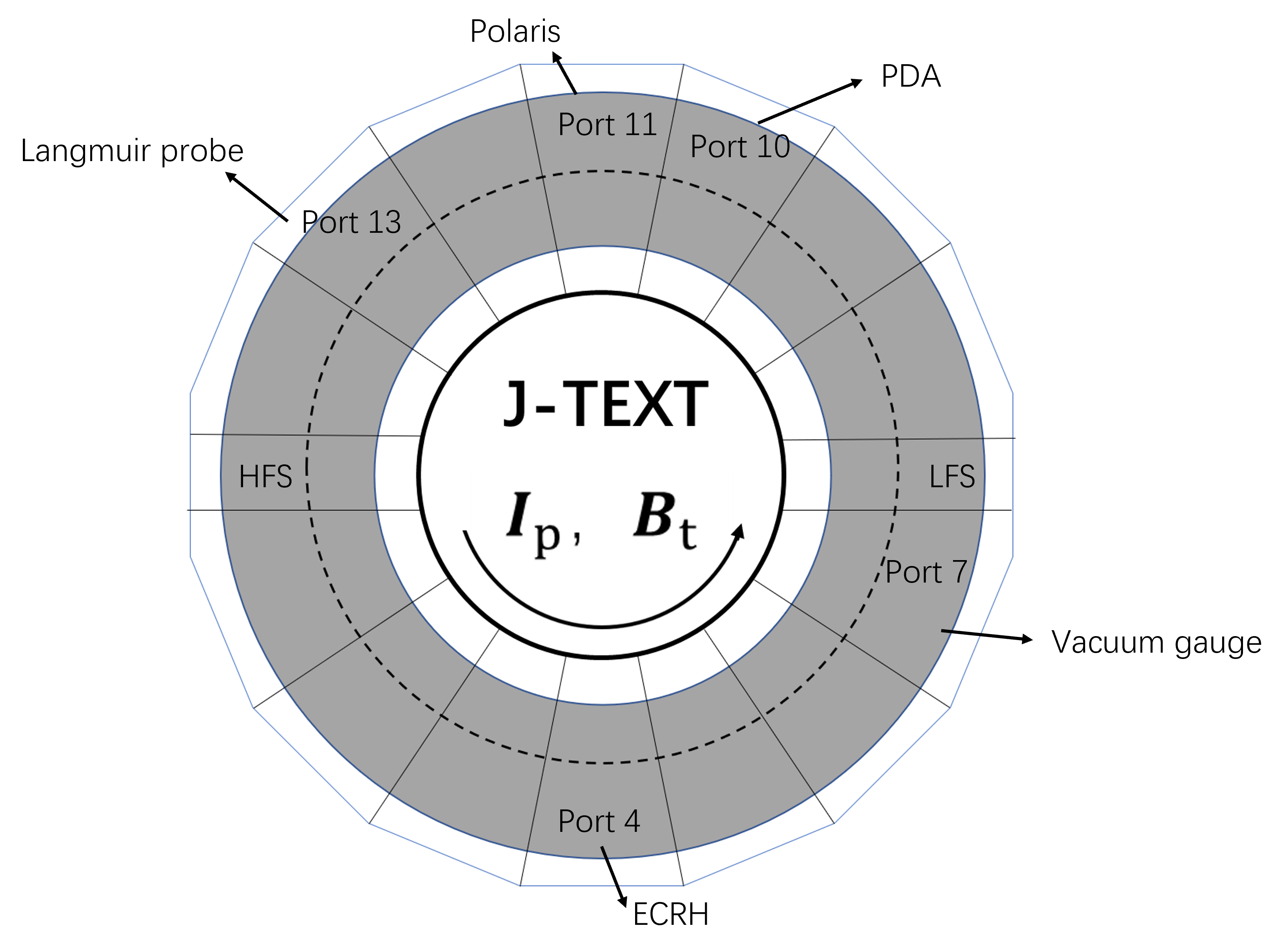}
	\caption{Top view of the relevant diagnostics on J-TEXT utilized in this study.}
	\label{fig:J-TEXTtopview}
\end{figure}


\newpage
\begin{figure}[htbp]
	\centering
	\includegraphics[width=1.2\linewidth]{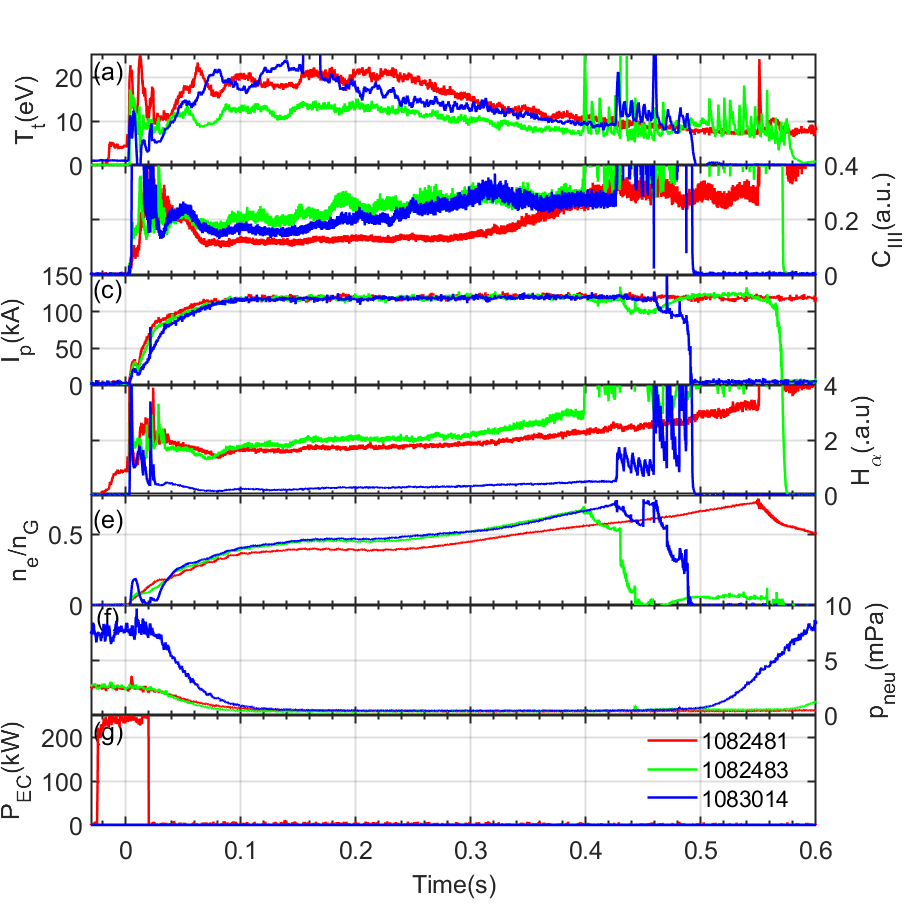}
	\caption{{Time histories of key parameters in} density limit
    experiments {with various} ECRH power and  pre-filled gas
    pressure in the start-up phase.}
	\label{fig:481_483_014}
\end{figure}

\newpage
\begin{figure}[htbp]
	\centering
	\includegraphics[width=1.2\linewidth]{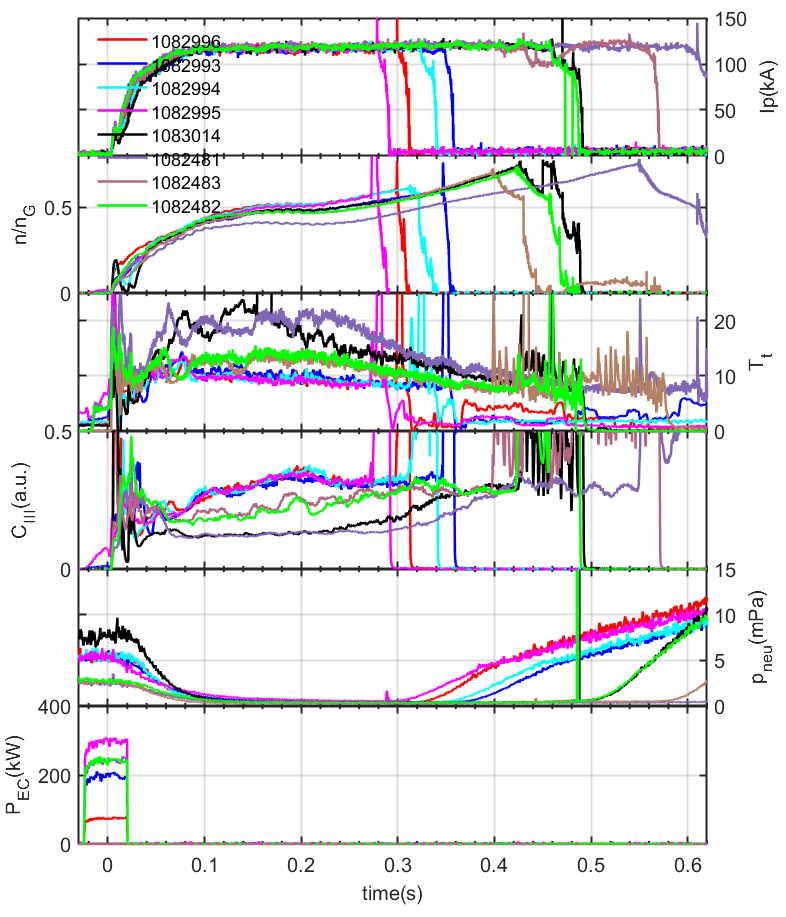}
	\caption{{Time histories of key parameters in} density limit
		experiments {with various} ECRH power and  pre-filled gas
		pressure in the start-up phase and the same plasma current.}
	\label{fig:moredata}
\end{figure}

\newpage
\begin{figure}[htbp]
	\centering
	\includegraphics[width=\linewidth]{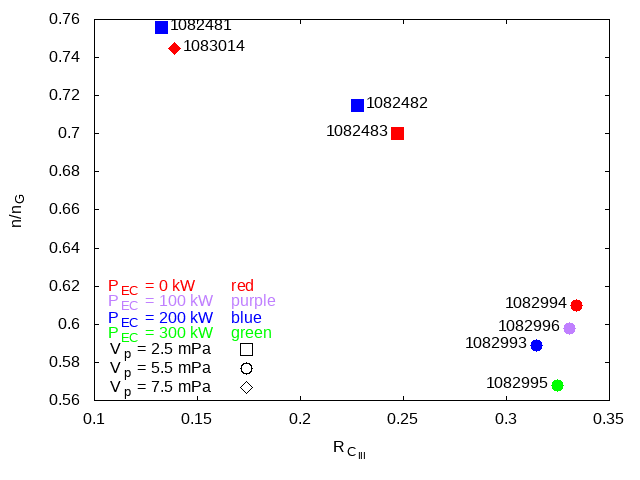}
	\caption{The density limit and {the corresponding}
     {radiation power} $R_{C_{III}}$ {measured from experiments for varying ECRH
      power and pre-filled gas pressure in the start-up phase}.}
	\label{fig:nlimit_CIII}
\end{figure}

\newpage
\begin{figure}[htbp]
	\centering
	\includegraphics[width=\linewidth]{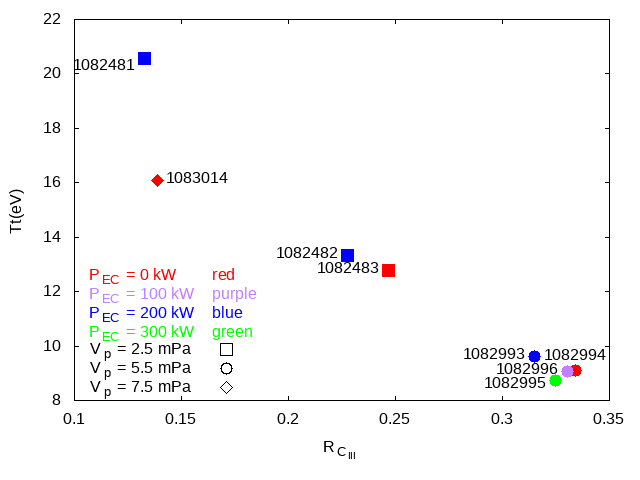}
  \caption{The plasma target region temperature $T_{\text{t}}$ and the corresponding
    impurity radiation power $R_{C_{III}}$ {measured from experiments for
      varying ECRH power and pre-filled gas pressure in the start-up phase}.}
	\label{fig:nlimit_Te}
\end{figure}

\newpage
\begin{figure}[htbp]
	\centering
	\includegraphics[width=\linewidth]{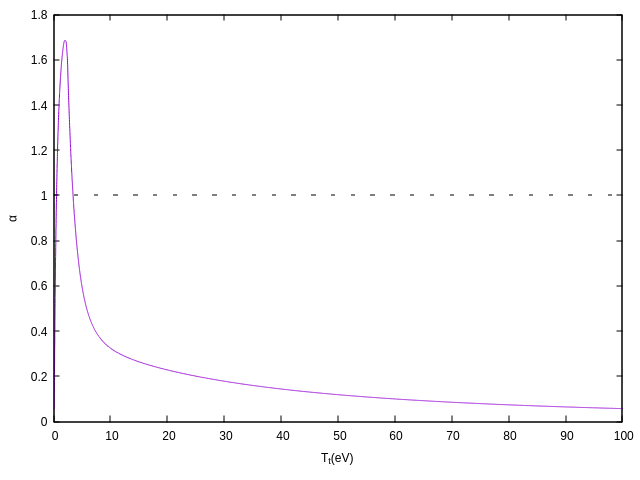}
	\caption{{The coefficient $\alpha$  {in Eq.~$\left(
				\ref{eq:alpha} \right)$ of the PWSO 0D model} as a function of the target
			region plasma temperature  {for a fixed plasma density}.}}
	\label{fig:alpha-Tt}
\end{figure}

\newpage
\begin{figure}[htbp]
	\centering
  \includegraphics[width=\linewidth]{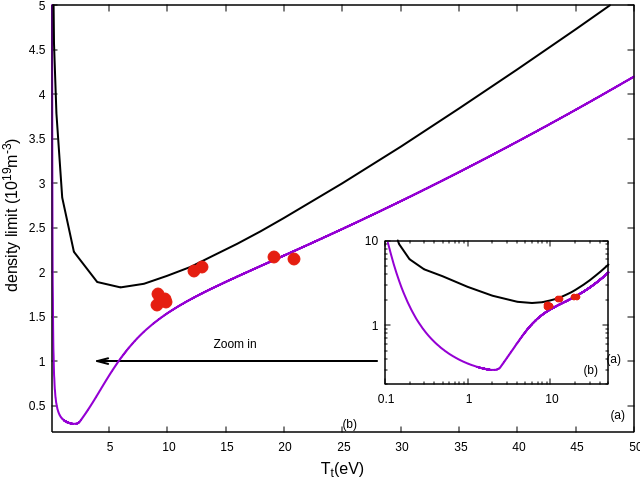}
	\caption{{The density limits as functions of the target region plasma
      temperature $T_{t}$  using (a): linear and (b): logarithmic coordinates}
    as predicted from  {the} PWSO 0D (purple solid line) and 1D
    (black line) models in comparison with the experimental
    data (red circular symbol).}
	\label{fig:denlim-Tt}
\end{figure}

\newpage
\begin{figure}[htbp]
	\centering
	\includegraphics[width=\linewidth]{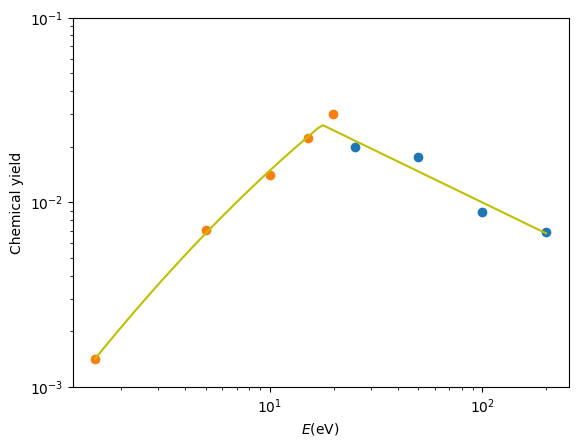}
	\caption{The fitted chemical sputtering yield function (solid line) and the
    original data from \cite{Stangeby2000} (blue points) and \cite{Joachim2007}
    (orange points).} 
	\label{fig:chem_sput}
\end{figure}

\end{document}